\documentclass[conference, 9pt]{IEEEtran}

\usepackage{mathptmx} 
\usepackage{xcolor}
\usepackage{graphicx}
\usepackage{listings}
\usepackage{framed}
\usepackage{fancyhdr}
\usepackage{tabularx}
\usepackage{adjustbox}
\usepackage[normalem]{ulem}
\usepackage[hyphens]{url}
\usepackage[sort,nocompress]{cite}
\usepackage[keeplastbox]{flushend}
\usepackage{booktabs}
\usepackage{amsmath}
\usepackage{amsfonts}
\usepackage{subcaption}

\definecolor{vgreen}{RGB}{104,180,104}
\definecolor{vblue}{RGB}{49,49,255}
\definecolor{vorange}{RGB}{255,143,102}

\lstdefinestyle{verilog-style}
{
    language=Verilog,
    basicstyle=\small\ttfamily,
    keywordstyle=\color{black},
    identifierstyle=\color{black},
    commentstyle=\color{black},
    numbers=left,
    basicstyle=\small,
    numberstyle=\tiny\color{black},
    numbersep=10pt,
    tabsize=8,
    moredelim=*[s][\colorIndex]{[}{]},
    literate=*{:}{:}1
}

\lstdefinestyle{C-style}
{
    language=C,
    basicstyle=\small\ttfamily,
    keywordstyle=\color{black},
    identifierstyle=\color{black},
    commentstyle=\color{black},
    numbers=left,
    numberstyle=\tiny\color{black},
    numbersep=10pt,
    tabsize=8,
    moredelim=*[s][\colorIndex]{[}{]},
    literate=*{:}{:}1
}

\makeatletter
\newcommand*\@lbracket{[}
\newcommand*\@rbracket{]}
\newcommand*\@colon{:}
\newcommand*\colorIndex{%
    \edef\@temp{\the\lst@token}%
    \ifx\@temp\@lbracket \color{black}%
    \else\ifx\@temp\@rbracket \color{black}%
    \else\ifx\@temp\@colon \color{black}%
    \else \color{black}%
    \fi\fi\fi
}
\makeatother

\usepackage{titlesec}
\titleformat{\paragraph}
{\normalfont\normalsize\itshape}{\theparagraph}{1em}{}
\titlespacing*{\paragraph}
{0pt}{3.25ex plus 1ex minus .2ex}{1.5ex plus .2ex}

\begin{document}

  \title{Countering the Path Explosion Problem in the Symbolic Execution of
    Hardware Designs\\
\author{\IEEEauthorblockN{Kaki Ryan} 
University of North Carolina \\
Chapel Hill, NC, USA
\and
\IEEEauthorblockN{Cynthia Sturton} 
University of North Carolina \\
Chapel Hill, NC, USA
}
}

\maketitle

%
%
%
%
%
\begin{abstract}


Symbolic execution is a powerful verification tool for hardware designs, but
suffers from the path explosion problem. We introduce a new approach, piecewise
composition, which leverages the modular structure of hardware to transfer the
work of path exploration to SMT solvers. We present a symbolic execution engine
implementing the technique. The engine operates directly over register transfer
level (RTL) Verilog designs without requiring translation to a netlist or
software simulation. In our evaluation, piecewise composition reduces the number of paths explored by an order of magnitude and reduces the runtime by 97\%. 
Using 84 properties from the literature we find
assertion violations in 5 open-source designs including an SoC and CPU.


\end{abstract}


\section{Introduction}

The verification of hardware designs is a key activity for ensuring the
correctness and security of a design early in the
hardware lifecycle. Current best practice includes assertion-based verification
(ABV) \cite{Coelho2004}, which has simulation-based testing as the underlying means of verification, and
formal verification techniques, an umbrella term encompassing many techniques
with the goal of proving a given property of a design. One technique in
particular that has gained recent attention, especially in security verification
applications, is symbolic
execution \cite{Shen2018SymbolicEB} \cite{zhang2018end} \cite{meng2021rtlcontest}.

Symbolic execution generalizes testing by
replacing input values with symbols, where each symbol represents the set of
possible values of the input parameter. A symbolic execution engine drives
symbolic execution using the semantics of the program's
language, but updated to include symbols. As execution proceeds the symbols are used in
place of literal values.
The result of symbolically executing a design for one clock cycle
is a tree of paths, each one associated with a unique \emph{path
condition} that describes the conditions satisfied by branches taken along the
path. If any path is found to violate a given assertion, then the associated path condition
acts as a precise description of the inputs that will drive
(concrete) execution along the same path; concrete  values that satisfy
the path condition are a counter-example to the assertion.

Unfortunately, symbolic execution suffers from the path explosion problem --
each path through a design is explored separately and the number of paths grows
exponentially with the number of branch points, or control flow statements, in the design.
Prior work has sought to avoid the path explosion
problem by combining symbolic execution with model checking~\cite{knox},
concrete execution traces~\cite{meng2021rtlcontest}, or by limiting the use to small
designs \cite{mukherjee2015hardware}.

We introduce \emph{piecewise composition}, a technique that leverages the structure of hardware designs to transfer
the problem to the domain of satisfiability modulo theories (SMT) solving so that the number of
paths to symbolically explore grows exponentially with only the number of branch
points in any one \texttt{always} block, and linear in the number of
\texttt{always} blocks in the design. 
In this way we reap the benefits of recent advances in SMT, while
maintaining the usability of having individual path information at the
register-transfer level (RTL).

Symbolic execution is closely related to symbolic simulation \cite{voss2} \cite{knox} \cite{sby}. In
both, concrete input values are replaced with symbolic values, representing
any possible value, and how the symbolic values propagate through the design is
tracked. However, there is a key difference. 
In symbolic simulation, the analysis is centered around dataflow. At
the end of a simulation run, each signal may hold the value true, false, or
a boolean expression characterizing the entire
circuit that drives that particular signal. Where there are control points in
the circuit, they are expressed as ITE statements in the boolean
expression. In symbolic execution, the analysis is centered around
control flow. At the end of one iteration, each signal is characterized by an
expression in first-order logic that characterizes the particular path taken through
the Verilog RTL. In addition there is a path
condition that represents the conditions under which execution would follow the
particular path through the design. 

There is a trade-off to be made between the complexity of queries sent to the
SMT solver (symbolic simulation) and the number of paths to explore (symbolic
execution). With piecewise composition, we examine a new point in the design
space, reducing the number of paths to explore to a tractable amount, while
still keeping SMT queries simple enough for modern solvers. The result is a
symbolic execution engine that can handle large designs and operate directly
over Verilog at the register-transfer level. 

Piecewise composition works by recognizing that independent parts of a design do
not need to be re-explored, once per root-to-leaf path. The algorithm symbolically explores each
 independent block of Verilog once, without consideration of the remaining
blocks, producing a set of symbolic execution trees. To reconstruct full root-to-leaf
paths, whether for finding assertion failures, describing how information flows through a design,
or to generate testcases, we can use SMT queries to combine the
independently explored path fragments.


Perhaps surprisingly, we show that for a design with $N$ always blocks, each
with at most $b$ binary branch points, symbolic execution of the design for a
single clock cycle requires symbolically executing $O(2^bN)$ paths, instead of
the $O(2^{bN})$ paths typical of symbolic execution. The number of paths to
explore grows exponentially with only the number of branch points in any one
independent block, and linearly with the number of blocks.

We apply piecewise composition to symbolically explore five open-source designs,
including SoC and
CPU designs, to find assertion violations in the design. Using 84 assertions from
the literature, we find that on average, piecewise
composition reduces runtime by 97\% compared to conventional symbolic
execution techniques without loss of efficacy.

This paper presents the following contributions: (1) Introduction and definition of \emph{piecewise
composition}, a technique that leverages the modular nature of hardware designs
to counter the path explosion problem in symbolic execution. (2) Design and implementation
of a symbolic execution engine for Verilog RTL using piecewise
composition. (3) Evaluation of piecewise composition and our implementation on
five open-source SoC and CPU designs. 

\section{Background}

We provide a review of the general techniques of symbolic execution and
SMT solving, and describe key aspects of
the Verilog hardware description language.

\subsection{Symbolic Execution}


In symbolic execution~\cite{king1976symbolic}, concrete literals are replaced
with symbolic values: input values are made symbolic and a symbolic execution engine
``executes'' the design, keeping track of the current execution state at
each line of code. The execution state has two main components: the symbolic
store $\sigma$ and the path condition $\pi$:
\begin{enumerate}
    \item $\sigma$, the symbolic store maintains mappings between program variables and symbolic expressions.
    \item $\pi$, the path condition is a boolean formula over symbolic
      expressions describing the conditions satisfied by branches taken along
      the current path. The path condition is always initialized to
      \texttt{True}.
\end{enumerate}

As the symbolic execution engine executes each line of code, using symbols in
place of literal values wherever they appear, the engine updates the symbolic
state. When a branching statement with condition $b$ is reached, the path condition $\pi$ is checked. If $\pi
\rightarrow b$, the \texttt{then} branch is taken. If $\pi \rightarrow \neg b$,
the \texttt{else} branch, if present, is taken. If neither implication holds,
then both branches must be explored in turn, forking the current path into two
separate paths to explore. To explore the path following the \texttt{then} branch, the path condition $\pi$ is updated:
$\pi:=\pi \wedge b$. To explore the path following the \texttt{else} branch, the
path condition is updated: $\pi:= \pi \wedge \neg b$. At each branch point, the
number of paths to explore doubles. This is the path explosion problem,
and typically heuristics or merging strategies~\cite{krishnamoorthy2010path,davidson2013fie,knox} are used to guide the exploration
to maximize coverage or depth.  

The complete exploration of a hardware design corresponds to a single clock
cycle of execution. Hardware executes
continuously and latent security vulnerabilities may only become clear many
clock cycles after the initial state. This requires symbolically executing the
design for multiple clock cycles, adding to the path explosion problem.

Importantly, our symbolic execution engine operates directly over the Verilog RTL
without translating to C or compiling down to the netlist. This allows
for greater human-readability of any found assertion violations -- the path
taken and the constraints over inputs will be directly traceable through the RTL design.
Our symbolic execution engine is cycle accurate. We assume no combinational latches, no
asynchronous resets, and always blocks are conditioned on input clocks. These
assumptions are in keeping with prior work in this area~\cite{knox}.

\subsection{Constraint Solving}

Boolean satisfiability (SAT) solvers are decision procedures that take in a
propositional formula and determine if there is an assignment of boolean values
that will make the formula evaluate to true. SAT modulo theories (SMT) solvers
generalize SAT solving by supporting
theories that are more expressive and can capture more functionality than simple
atomic propositions. Examples include supporting array operations and supporting
linear arithmetic. SMT solvers have become quite powerful and are able 
to handle expressions with hundreds of variables. However, SMT solving is still an NP-complete problem. 

SMT solvers are crucial to symbolic execution both in checking feasibility of paths as
the engine progresses, and in generating assignments to
symbolic variables for a found path, e.g., to produce a test-case. Some of the most widely used solvers are STP~\cite{STP}
(used in KLEE~\cite{cadar2008klee}) and Z3~\cite{de2008z3} (used in Mayhem~\cite{cha2012mayhem} and angr~\cite{kim2017angr}). 

\subsection{Verilog}
Verilog is a hardware description language and is the industry standard for developing real-world computer systems. A basic unit of design in Verilog is a module. Modules often contain
other modules, making the design hierarchical. A module combines multiple sub-modules by making the output signals of one module connect to the input signals of a second
module, with connection wires and registers in between. Verilog has several
constructs that allow data to flow differently than a sequential-only software
program would.  For example, multiple modules composed in parallel are truly
executing in parallel. Within a module, an $\texttt{always}$ block is
used to define a set of events that only happen under certain conditions. For instance, assignment statements that are only to 
be executed at a clock's rising or falling edge. The statements within a sequential $\texttt{always}$ block are all
executed in parallel, and two different $\texttt{always}$ blocks operate in parallel.

\section{Piecewise Composition}



Symbolic execution has two main costs: the time required to simulate execution and
update the symbolic store for each line of code, and the time required to 
determine whether both branches are feasible at each branch point in a path. Both parts are costly, but recent
advances in SMT solving have cut the cost to deciding branch feasibility,
whereas the costs to symbolically execute an instruction have remained
relatively stable.

In conventional symbolic execution, each line of code is potentially visited multiple
times, once for each path explored.
Our approach is to aggressively decompose the design into independent blocks, symbolically
explore each block once, then use an SMT solver to compose path conditions
and symbolic stores from each block. This strategy is made possible by the
inherent modular nature of hardware designs, and lets us leverage the relative speed of modern
SMT solvers compared to the cost of symbolically executing lines
of code.

The result is
a complete (logical) tree of paths through the design. While the
number of paths in the full tree is exponential in the number of branches in the
design, the symbolic execution engine
explores a number of paths exponential in only the number of branches in any single
independent block and polynomial in the number of blocks. 



%

\subsection{Motivating Example}

\begin{figure}[h]
  \centering
  \begin{framed}
    \begin{lstlisting}[style={verilog-style}]
always @ (posedge clk) begin
   if (g0)
      x <= inpA; //inpA is an input signal
   else
      x <= 0;
end

always @ (posedge clk) begin
   if (g1)
      y <= inpB; //inpB is an input signal
   else
      y <= 0;
end
   
    \end{lstlisting}
  \end{framed}
  \caption{Verilog code example with two branches.}
  \label{fig:twobranchpoints}
\end{figure}

  \begin{figure}[h]
  \centering
  \begin{subfigure}[b]{0.18\textwidth}
    \includegraphics[width=\textwidth,trim = 0 0 560 75, clip]{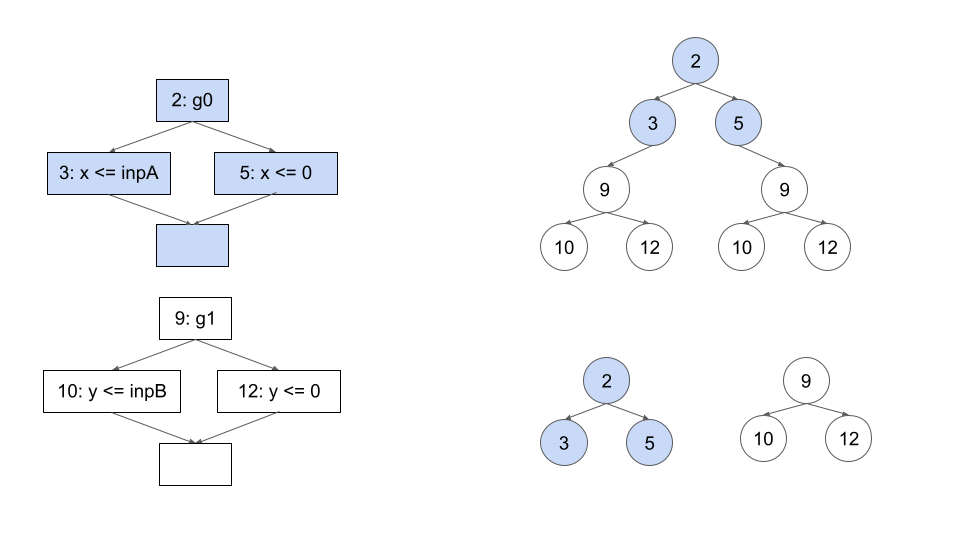}
    \caption{Control flow graph}
    \label{fig:cfg}
  \end{subfigure}
    \begin{subfigure}[b]{0.25\textwidth}
    \includegraphics[width=\textwidth,trim = 500 200 25 30, clip]{figures/toyexample}
    \caption{Full tree of paths}
    \label{fig:tree}
  \end{subfigure}
  \begin{subfigure}[b]{0.35\textwidth}
    \vspace{2pc}
    \includegraphics[scale = 0.40, trim = 520 50 45 350, clip]{figures/toyexample}
    \caption{Trees of paths to be independently explored under piecewise
      composition}
    \label{fig:decomposedtree}
  \end{subfigure}
  \caption{Piecewise Composition}
    \label{fig:piecewisecomp}
\end{figure}

Figure~\ref{fig:twobranchpoints} shows a code snippet with two always blocks
and branch points at
lines 2 and 9. The corresponding control flow
graph with an arbitrary ordering of the always blocks is given in Figure~\ref{fig:cfg}, and the tree of paths through the design
is given in Figure~\ref{fig:tree}. With conventional symbolic execution, each of the four
root-to-leaf paths in Figure~\ref{fig:tree} is
symbolically executed. This is the strategy taken by current approaches (e.g.,~\cite{zhang2018end,fowze2022eisec}) that translate
a design into a C++ representation and then use the KLEE
symbolic execution engine. But, the two subtrees rooted at a node labeled 9 represent
repeated work. For each subtree, the symbolic execution engine was exploring the
paths through the same block of code: the second \texttt{if-else} block starting at line 9.

The branching condition and assignments in lines
2--5 are independent of the branching condition and assignments in lines
9--12. Regardless of which path is taken at the first branch (line 2), the
symbolic execution starting at the second branch point (line 9) will produce the same
sub-tree. The feasibility of paths of the second condition will be the same, and
updates to the symbolic state will be the same. For example, let inputs
$\mathtt{inpA}$ and $\mathtt{inpB}$ be initialized with symbolic values $\alpha$
and $\beta$, respectively, and assume the gating signals $\mathtt{g0}$ and
$\mathtt{g1}$ have symbolic values $\gamma_0$ and $\gamma_1$ at this point in the
code. For the path in which both branches are taken (nodes
$\langle\mathtt{2, 3, 9, 10}\rangle$ in the symbolic execution tree), the
symbolic store at the end of the path would be $\sigma = \{x := \alpha, y := \beta\}$
with path condition $\pi = \gamma_0 \wedge \gamma_1$. For the path in which the first
branch is not taken, but the second one is ($\langle\mathtt{2, 5, 9, 10}\rangle$),
the symbolic store at the end of the path would be $\sigma = \{x := 0, y:= \beta\}$
and the path condition is $\pi = \neg \gamma_0 \wedge \gamma_1$. In both paths, the
updates to $y$ are the same, despite the different updates to $x$.

\subsection{Piecewise Composition}

With piecewise composition, the
symbolic execution engine explores independent blocks of the RTL
separately, producing independent trees of path fragments. In the above example, piecewise composition
results in the two trees produced in Figure~\ref{fig:decomposedtree}. The symbolic
execution engine now explores the second \texttt{if-else} block only once. Continuing with our example, piecewise composition will separately
explore the two always blocks, producing the following four path fragments
(labeled $a$ through $d$) with
associated path conditions and (partial) symbolic stores:
\begin{align*}
  \langle\mathtt{2, 3}\rangle_a: \qquad & \sigma_a = \{x := \alpha\}, \pi_a = \gamma_0 \\
  \langle\mathtt{2, 5}\rangle_b:  \qquad& \sigma_b = \{x := 0\}, \pi_b = \neg \gamma_0 \\
  \langle\mathtt{9, 10}\rangle_c:  \qquad& \sigma_c = \{y := \beta\}, \pi_c = \gamma_1 \\
  \langle\mathtt{9, 12}\rangle_d:  \qquad& \sigma_d = \{y := 0\}, \pi_d = \neg \gamma_1 \\
\end{align*}

To
find full paths through the design and to successfully find assertion violations,
all realizable combinations of path fragments are composed with the help of an
SMT solver. For example, to realize path $\langle\mathtt{2,5,9,10}\rangle$, the
symbolic execution engine queries the SMT solver to find whether the two path
fragments, $\langle\mathtt{2, 5}\rangle$ and $\langle\mathtt{9, 10}\rangle$ can
be joined: $\mathtt{isSAT}(x = 0 \wedge \neg \gamma_0 \wedge y = \beta \wedge
\gamma_1)$. In this simple example, all four combinations of path fragments are
possible, but in general that will not always be the case. If, for example, the
gating signals $\mathtt{g0}$ and $\mathtt{g1}$ both took their values from the
same input, and were therefore constrained to always have the same value, the
SMT solver would find that paths $\langle\mathtt{2, 5, 9, 10}\rangle$ and
$\langle\mathtt{2, 3, 9, 12}\rangle$ were unrealizable.

An important result of piecewise composition is that it is sound: all composed paths that are realizable correlate to replayable paths
through the full design. The satisfying solutions returned by the SMT solver
can be provided as inputs to the design in the starting reset state and will
result in execution (either in simulation or running on an FPGA) following the corresponding path.

\subsection{Comparison with Backtracking and Caching}
Piecewise composition shares some similarities with backtracking and caching, two techniques often used in
software symbolic execution engines (e.g., KLEE~\cite{cadar2008klee}, Angr~\cite{kim2017angr}). But, there are key differences. Backtracking reduces
repeated work by maintaining state at each point in a path and allowing two
paths with a shared prefix to reuse the saved state. For example, If path $\langle\mathtt{2, 3, 9, 10}\rangle$ 
has been explored, then when the engine explores path
$\langle\mathtt{2, 3, 9, 12}\rangle$, backtracking allows the engine to reuse the saved state at point 9 and
continue exploration from there. Backtracking prevents
re-exploring path $\langle\mathtt{2, 3}\rangle$ for each of
$\langle\mathtt{9, 10}\rangle$ and $\langle\mathtt{9, 12}\rangle$;
piecewise composition also prevents this re-exploration. However, with backtracking, paths
$\langle\mathtt{9, 10}\rangle$ and $\langle\mathtt{9, 12}\rangle$ will be
re-explored to create the paths starting with prefix $\langle\mathtt{2, 5}\rangle$; this
re-exploration is prevented by piecewise composition.


Caching queries reduces
the time spent in the SMT solver by reusing the results from prior
queries. Caching queries is a technique orthogonal to piecewise
composition. Using the two techniques together could further reduce runtime.

\section{A Symbolic Execution Engine with Piecewise Composition}
Mechanically, the symbolic execution engine achieves piecewise composition by decomposing a design into partitions: one partition to contain all
combinational logic in the design, one partition for all register declarations,
and a set of $N$ partitions, one per
\texttt{always} block, to handle the sequential logic in the
design. Each partition is fully symbolically explored once per clock cycle, with the exception
of the combinational logic partition, discussed next in Section~\ref{sec:combinationallogic}.

Each of the $N$ sequential \texttt{always} block partitions are explored
independently of the other \texttt{always} blocks, and the
exploration produces a set of path
fragments. The complete exploration of the full design produces $N$ sets, one
per always block. The set of full root-to-leaf symbolic execution paths through the design
is formed by taking the cross-product of the $N$ sets of path fragments. The SMT
solver is used to ensure only those combinations that are sound -- that
correspond to true paths through the design -- are kept.


\subsection{Combinational Logic}
\label{sec:combinationallogic}
The symbolic execution engine will check for any combinational latches, and
  if any appear, will exit with an error.  Otherwise, the tool symbolically executes each combinational logic statement in the partition and then begins to symbolically
  execute the control flow paths through each \texttt{always} block. As each block is executed the engine keeps track of a dirty bit for each signal, which gets set to 1 when the
signal is updated. Once a path has been completed,
every \texttt{assign} statement in the combinational logic partition for which the right-hand side involves a dirty
signal is re-evaluated, in order. During this re-evaluation, the engine continues to
track when signals become dirty. Additionally, conditional continuous assignments
using the ternary operator are characterized as ITE statements rather than
branch points when they are evaluated during this phase.

\subsection{Sequential Logic}

Each sequential \texttt{always} block is explored independently.
This approach is sound if
the \texttt{always} blocks are truly independent -- the path condition and
symbolic state of the various paths through one block are the same regardless of
the paths taken through other blocks. In the following we
discuss the issue of independence in more detail. Consider two sequential \texttt{always}
blocks, $B_0$ and $B_1$, both blocks are triggered on the rising edge of the input clock
signal. (The choice of which clock edge is used is irrelevant, but it is assumed
that both blocks trigger on the same edge of the same clock signal.)

\textbf{Independence} In the simplest case, none of the variables that appear in $B_0$ appear in
$B_1$. The two blocks are independent and, within a single clock cycle, the execution of one block has no
bearing on the execution of the second block. (Recall that two sequential \texttt{always}
blocks can be thought of as executing in parallel.) The two blocks can be explored
separately and their paths can be composed in any order. This case is rare, however, as an
input reset signal typically appears in all or most blocks.

\textbf{Read-read dependence} In the next case, the same variable may appear in
a branch condition or right-hand side of an assignment in both
$B_0$ and $B_1$. The two blocks can still be explored
separately and their paths can be composed in any order. However, some
combinations of paths may not be feasible, as variables that appear in branch conditions in both blocks, say $b_0$ and $b_1$, respectively, will preclude the combination of paths from
$B_0$ in which $b_0$ holds with paths from $B_1$ in which $b_1$ holds when $b_0
\wedge b_1$ is unsatisfiable. 

\textbf{Read-write dependence} In the next case, a variable may appear in a
branch condition or on the right-hand side of an assignment in
$B_0$ and on the left-hand side
of an assignment in $B_1$. When non-blocking assignments are used,
updates to variables in $B_1$ take effect in the next clock-cycle, whereas
reads and conditional branches in $B_0$ are using variable values as set
in the previous clock cycle. The symbolic execution engine keeps the
appropriate value and there is no conflict. The two blocks can be
  explored separately and their paths can be composed in any order. Our tool
  does not support the use of blocking assignments within sequential
  \texttt{always} blocks.

  \textbf{Write-write dependence} In the final case, variables appear on the left-hand side of an assignment in both
\texttt{always} blocks. This violates best practice in Verilog design. The
symbolic execution engine will check for any instances of write-write dependence,
and if any appear will exit with an error.

\subsection{Multiple Clock Cycles}
The symbolic execution of a hardware design for one clock cycle produces a tree
where the root node represents the design in the reset state or initial state,
and each each leaf node corresponds to a ``next-state'' at the clock-cycle
boundary. To explore a design for two clock cycles would mean exploring the
design again, once for each leaf node of the tree constructed during clock cycle
0. This time exploration would begin not at the reset state, but rather at the
state indicated by the given leaf node of the prior cycle. At the end of each
single clock-cycle exploration, there is a merging process in which all of the
input signals and combinational wires are given fresh symbols while stateful
elements carry over their expressions from the previous cycle.



\subsection{Further Optimizations}
\label{sec:optimizations}
\subsubsection{Repeat Submodules}
When the modules are duplicate instantiations of the same module there is room for reduction in the total search space. The main idea of the optimization is similar in spirit to piecewise composition in that we explore each submodule once for each path. Then instead of re-exploring again for each repeat
instantiation, we can merge in the symbolic store and path condition for the given root-to-leaf path using SMT queries.
\subsubsection{Cone of Influence Analysis}
 This optimization prunes the exploration space at the block level. The
 symbolic execution engine will read in the expressions supplied in the assertions, perform a dependency analysis over the signals in the assertions
 and then complete an AST traversal to determine which blocks read from or write
 to the signals of interest or their dependencies. 
 After this initial pass, the engine will only explore blocks that involve
 the signals of interest or their dependencies. 

\section{Implementation}
Our symbolic execution engine is built in python 3.8 and implements the Verilog
semantics according to the IEEE 1364-2005 standard. We use the pyVerilog library to
build the Verilog AST and we use networkX to manage graph search and
traversal. We use the Z3 python API for SMT solving. The symbolic execution
engine takes in a design written in Verilog, 
including the assertions written according to the SystemVerilog 1800-2017 standard,
and outputs replayable counterexamples.

\section{Evaluation}

We evaluate our implementation over five open-source designs, including CPU and SoC designs,
to study its viability as a platform for the verification of hardware designs. Our evaluation
considers the following questions: 1) How well does piecewise composition
counter the path explosion problem? 2) What effect do piecewise
composition and the optimizations described in Section~\ref{sec:optimizations}
have on performance? 3) Can the symbolic execution engine produce assertion
violations with replayable counter-examples for buggy and vulnerable designs?  





\subsection{Dataset and Experimental Setup}


We collected five designs and 84 security critical assertions from several
sources. The first three designs and associated assertions came from the
Security Property/Rule Database available on TrustHub~\cite{TrustHub1,
  TrustHub2}: an enhanced version of the Serial Peripheral Interface
available on Motorola's MC68HC11 family of CPUs; openMSP430, a synthesizable 16-bit microcontroller core compatible with Texas Instruments' MSP430
microcontroller family; and a CrypTech True Random Number Generator (TRNG). For each of these designs, the database included 9, 2, and 2
security properties, respectively.

  The fourth design is the PULPissimo SoC used in a recent Hack@DAC competition \cite{HACK}. The design is buggy; some of the bugs were inserted manually by the organizers of
  the competition and others were native to the
  design~\cite{hardfails2019}. Using the English description of the properties,
  as well as the walkthrough of the test-case generation in the RTL-ConTest
  paper~\cite{meng2021rtlcontest} we developed 26 assertions for use with our tool. 

The fifth design is the OR1200 processor core. We collected 30 security-critical bugs from two prior papers, SPECS~\cite{hicks2015specs} and SCIFinder~\cite{zhang2017scifinder} and 70 security assertions from SPECS~\cite{hicks2015specs}, Security Checkers~\cite{bilzor2011security}, SCIFinder~\cite{zhang2017scifinder}, and Transys~\cite{zhang2020transys}. 

The experiments are performed on a machine with an Intel Xeon E5-2620 V3 12-core
CPU (2.40GHz, a dual-socket server) and 62G of available RAM. 
 \subsection{Mitigation of Path Explosion}

\begin{table*}[htb]
  \small
  \centering
    \begin{tabular}{lrrrrrr}
    \toprule
    Design & \multicolumn{2}{c}{Baseline} &
        \multicolumn{2}{c}{Piecewise Composition} & \multicolumn{2}{c}{Percent Decrease} \\
    \cmidrule(l){2-7}
    & \multicolumn{1}{c}{LoC} & \multicolumn{1}{c}{branch points} &
    \multicolumn{1}{c}{LoC} & \multicolumn{1}{c}{branch points} &
    \multicolumn{1}{c}{LoC} & \multicolumn{1}{c}{branch points} \\
    & \multicolumn{1}{c}{explored} & \multicolumn{1}{c}{explored} &
    \multicolumn{1}{c}{explored} & \multicolumn{1}{c}{explored} &
    \multicolumn{1}{c}{explored} & \multicolumn{1}{c}{explored} \\

    \midrule
  OR1200 & 54018 & 7803 & 881 & 45 & 98\% & 99\% \\
  Hack@DAC & 493032 & 15093 & 3525 & 276  & 99\% & 98\% \\
  MC68HC11 SPI  & 2093 & 158 & 174 & 57 & 92\% & 64\%  \\
  openMSP430 & 15293 & 377 &  489 & 68 & 97\% & 82\% \\
  CrypTech TRNG & 8930 & 421  & 336 & 91 & 96\% & 78\% \\
    \bottomrule
  \end{tabular}
  \caption{Average Impact of Piecewise Composition on Path Explosion}
  \label{lst:pcsummary}
  \end{table*}

We evaluate how well piecewise composition mitigates the path
explosion problem. The number of paths extant in a design will not change, but the
amount of work the symbolic execution engine has to do to realize a complete
path is considerably smaller with piecewise composition. For each design we compare the average number of lines of code and
branch points visited to find an assertion violation both with and without
piecewise composition. Table~\ref{lst:pcsummary} has the results. Piecewise
composition reduces the number of lines of code visited (i.e., reduces redundant
visits to the same line of code) by 92\%--99\% and branch points visited by 64\%--99\%.
  
To gain a more complete picture, we
symbolically explore all paths through one design. We used the MC68HC11
SPI, which is small enough that symbolic execution without piecewise composition
can complete the exploration. This design has 13 \texttt{always} blocks and 1459
paths through the design. These results are reported in
Table~\ref{lst:SPIpaths}. Without piecewise composition more than 90k lines of
code and more than 7k branch points are explored. With piecewise
composition, the symbolic execution engine needs to explore roughly only 7\% of those
90k lines of code and only 4\% of those 7k branch points.

    \begin{table}[htb]
  \small
  \centering
  \begin{tabular}{p{1.2in}rrr}
    \toprule
    Configuration &  \multicolumn{1}{c}{LoC} & \multicolumn{1}{c}{branch points} & \multicolumn{1}{c}{paths} \\
     &   \multicolumn{1}{c}{explored} &  \multicolumn{1}{c}{explored} &  \multicolumn{1}{c}{completed} \\
    \midrule
  Baseline & 90706 & 7380  & 1459 \\
  Piecewise Composition & 6783 & 323  & 1459 \\
    \bottomrule
  \end{tabular}
  \caption{Full Exploration of MC68HC11 SPI Design}
  \label{lst:SPIpaths}
  \end{table}


    \subsection{Effects of Optimizations}

\begin{figure*}[htb]
  \begin{subfigure}[b]{0.5\textwidth}
    \includegraphics[width=\textwidth]{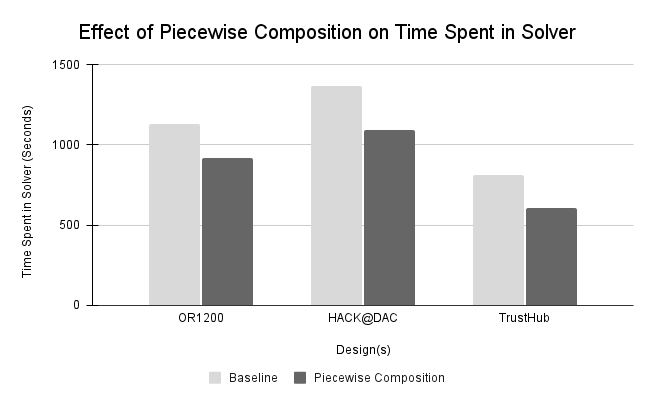}
    \caption{Solver Time}
    \label{fig:solvertime}
  \end{subfigure}
  \begin{subfigure}[b]{0.5\textwidth}
    \includegraphics[width=\textwidth]{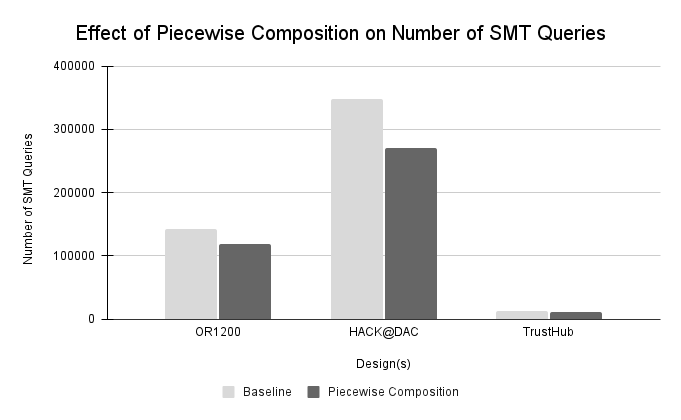}
    \caption{SMT Queries}
    \label{fig:queries}
  \end{subfigure}
  \caption{Effects of Optimizations on Solver Time and Number of SMT Queries}
  \label{tab:smtgraphs}
  \end{figure*}

Figure~\ref{tab:smtgraphs} shows the impact of piecewise composition on the
average number of SMT queries and average time spent in the SMT solver for each
design. One concern might be that the win in minimizing redundant explorations
comes at the expense of exploding SMT solver work. However, the opposite
occurs. Reducing the number of paths explored also reduces the number of queries
to the solver overall; remember that during exploration, the solver is queried
at each branch point. With piecewise
composition turned on there was an 18\% decrease, on average, in the number of
SMT queries and a 21\% decrease in the amount of time spent solving.

In Table~\ref{tab:optimizations} we measure runtime for four cases: \emph{Baseline}, with no optimizations enabled; \emph{P.C}, with piecewise composition enabled, \emph{Repeat}, with repeated modules explored only once; and
\emph{COI}, with cone-of-influence analysis completed before exploration. 
Each case is cumulative, for example, in the Repeat case
the Parallel optimization is enabled as well. For each design we take the
average when looking for each assertion. For all but the smallest design,
Baseline could not reliably complete exploration within 30 minutes at which point we
stopped searching. Table~\ref{tab:optimizations} shows the
results. For each case, we provide the absolute runtime in seconds and the
decrease in runtime compared to the prior case. Overall, the
optimizations decrease the engine's runtime by 95-99\%. 

\begin{table*}[htb]
\small
  \centering
    \begin{tabular}{lrrrrrrrr}
    \toprule
    Design & \multicolumn{1}{c}{Baseline} &
        \multicolumn{2}{c}{Piecewise} & \multicolumn{2}{c}{Redund} & \multicolumn{2}{c}{COI}  & \multicolumn{1}{c}{Overall} \\
    \cmidrule(l){2-9}
    & runtime & runtime & \% dec &
    runtime  & \% dec & runtime  & \% dec & \% dec \\
    & (sec) & (sec) &  &
    (sec) &   & (sec) &  &  \\
    \midrule
    OR1200 & timeout (1800) & 52.47 &  97.08\% & 37.56 & 21.31\% & 25.22 & 12.56\%  & 98.60\%\\
    Hack@DAC & timeout (1800) & 174.24 &  90.32\% & 121.94 & 28.34\% & 81.83 & 16.62\% & 95.45\% \\
    MC68HC11 & 962 & 17.53 &  98.18\% & 14.30 & 19.93\% & 0.07 & 99.19\% & 99.99\% \\
    openMSP430 & timeout (1800) & 37.65 & 97.91\%   & 23.14 & 38.55\% & 0.73 & 96.83\% & 99.96\%  \\
    CrypTech TRNG & timeout (1800) & 14.92 &  99.17\% & 12.08 & 19.15\% & 0.09 & 99.19\% & 99.99\%  \\
    \bottomrule
  \end{tabular}
  \caption{Average Effect of Optimizations on Runtime}
  \label{tab:optimizations}
  \end{table*}

\subsection{Finding Assertion Violations}

To evaluate the engine's ability to find assertion violations, we run a set of
experiments in which we have ground-truth knowledge of the (minimum) number of
violations in each design. From our dataset, we had ground truth for two of the designs: the
Hack@DAC SoC (31 bugs) and the OR1200 processor core (30 bugs). 

In these experiments, symbolic execution begins in the reset state with all input signals
made symbolic, and execution continues until an assertion violation is found.
 Table~\ref{tab:knownbugs} summarizes the results.  

The engine finds 24 of the 31 bugs in the Hack@DAC SoC. By way of comparison,
prior work reports finding 14 of the 31 bugs using concolic
execution~\cite{meng2021rtlcontest} and the organizers of Hack@DAC report
finding 6 and 15 bugs using the commercial tools Cadence SPV and Cadence FPV,
respectively~\cite{HACK}. 

The engine finds 27 of the 30 bugs in the OR1200. Of the three missed, one did not
have a property that covered it. The
second bug was related to the correct instruction being executed and would allow
an attacker to carry out an ROP exploit by early kernel exit. The third bug was
related to control flow and will incorrectly set the link register, which is
used when the processor returns from function calls.

All
counterexamples generated by our engine were successfully replayed in simulation
starting from the reset state using Vivado.


 \begin{table}[htb]
   \small
 \centering
 \begin{tabular}{lcp{.4in}p{.4in}p{.4in}}
    \toprule
     Design & \# Bugs & \# Bugs Found & Avg Time (sec) & Max Clock Cycles Taken \\
     \midrule
     Hack@DAC & 31 & 24 & 122 & 4 \\
     OR1200 & 31 & 27 & 34  & 5 \\
    \bottomrule
  \end{tabular}
   \caption{Known Bugs}
   \label{tab:knownbugs}
   \end{table}

\subsection{Comparison to Current State of the Art}

Coppelia is a tool that first translates Verilog to C and
then uses a conventional symbolic execution, but with search strategies
optimized for hardware designs~\cite{zhang2018end}. The authors report that most (62\%) of
exploits in their experiments are generated within 15 minutes, and several are
found within 2--4 hours. Symbolic execution with piecewise composition, by contrast, finds the same exploits in comparable
designs (including the OR1200) in under 2 minutes. 
The authors of RTLConTest~\cite{meng2021rtlcontest}, a concolic execution
engine, report that it takes around an hour and 40 minutes to complete on the PULpissimo SoC, which is a modified version of the HACK@DAC 2018 design that we used for our evaluation. Once they perform the concolic execution and generate the tests, it takes 10 seconds on average to produce a counterexample. For security properties targeting the same bugs, our tool performs the complete end-to-end symbolic execution workflow to generate counterexamples in 122 seconds, on average.

\section{Related Work}


\textbf{Model Checking}
Current practice in hardware validation involves a combination of
simulation-based testing and static analysis techniques like model
checking. Model checking can formally verify that a program satisfies a set of
specifications, given in the form of a temporal logic formula
\cite{modelchecking}. The general verification procedure is an exhaustive search
of the design space. In practice, bounded model checking is used in which the
verification explores the design for up to k time steps \cite{BMC}. Symbolic
Trajectory Evaluation is a lattice-based model checking technique developed and
used by Intel \cite{STE2}, and more recent work has brought this closer to the
word-level \cite{word-STE}. The goal of STE is to formally verify properties of
a sequential system over bounded-length trajectories using a modified form of
3-valued symbolic simulation \cite{STE}. SymbiYosys \cite{sby} is a model checking engine built on top of Yosys
that serves as an extension of Yosys \cite{yosys} existing property checking framework.

\textbf{Symbolic and Concolic Execution}

The use of symbolic execution and related techniques for RTL designs is gaining
traction. RTLConTest \cite{meng2021rtlcontest}  and the work by Witharana et al.~\cite{Witharana2021Concolic} are
examples of concolic testing engines developed for the security verification of
hardware designs. Coppelia \cite{zhang2018end} is a hardware-oriented backward
symbolic execution engine built on top of KLEE for RTL designs translated to
C++. EISec~\cite{fowze2022eisec} also uses KLEE, but for netlists translated to
C++. All demonstrate the power of symbolic execution as applied to hardware designs,
but all still struggle with the path explosion problem. Many of the techniques
developed in those papers can be combined with piecewise composition.


\textbf{Fuzzing}
Fuzzing has also been shown to be a useful technique for finding security vulnerabilities in SoCs and CPU designs. RFUZZ is a coverage-directed fuzz tester for circuits that presents a hardware specific coverage metric called \textit{mux control coverage} \cite{Laeufer2018Mux}. DifuzzRTL is an RTL fuzzing tool used to find unknown security bugs that measures coverage based on control registers rather than multiplexors’ control signals to improve efficiency and scalability \cite{Hur2021RTLfuzz}. A recently developed Hardware Fuzzing Pipeline translates the RTL to a software model to improve scalability in bug finding via fuzzing \cite{fuzzlikesw}.

\textbf{Information Flow Tracking}
In a hardware context, information flow refers to the transfer of information between different signals. Information flow tracking is a verification method that studies how information flows through a hardware design to make statements about security \cite{IFT1} \cite{IFT2}. A hardware design can be instrumented with tracking logic to capture timing \cite{ift2017timing} or data flow information. This method can provide strong security guarantees, for example, demonstrating leakage of secret key data to undesired output signals.

\section{Conclusion}

We have presented piecewise composition, a technique for countering the path
explosion problem in symbolic execution. We implemented a symbolic
execution engine using the technique and evaluated the engine on five
open-source designs. The engine reduces redundant
work by
98\%--99\% compared to conventional symbolic execution, improves overall
performance and successfully finds
assertion violations.  

\section{Acknowledgments}
This material is based upon work supported
by the National Science Foundation under Grant No. CNS-1816637, and by a Meta Security Research Award.

\end{document}